# Accounting the size distribution of HTS granules for the critical current density from magnetic measurements

D. M. Gokhfeld[1*], Yu. S. Gokhfeld[1]

[1] Kirensky Institute of Physics, Krasnoyarsk Scientific Center, Siberian Branch, Russian Academy of Sciences, 660036 Krasnoyarsk, Russia

[*] e-mail: gokhfeld@iph.krasn.ru

The determination of the critical current density from magnetic hysteresis loops is widely used to characterize and compare superconducting samples. Magnetic hysteresis loops for tapes and single crystals depend on both the critical current density and sample size. The latter sets the scale of supercurrent circulation. However, in polycrystalline high-temperature superconductors prepared by solid-phase synthesis or by sol-gel method, the magnetization is determined by the circulation of supercurrents in individual grains. The paper discusses the effect of the grain size distribution on the effective scale of current circulation. Log-normal and Weibull distributions are both considered as possible for grain sizes. The effective size for calculating the intragrain current density has been shown to be significantly larger than the average grain size.

## 1. Introduction

The critical current density $J_c$ is the most important parameter for characterizing a superconductor to its practical applicability at a given temperature [1]. Direct measurement of $J_c$ in high-temperature superconductors (HTSC) is very difficult due to its extremely high values, reaching $10^{12}$ A/m$^2$ [2]. A commonly accepted indirect approach to determining $J_c$ is through magnetic hysteresis loops. This method utilizes the ratio between $J_c(H)$ and $\Delta M(H) = M_{dec}(H) - M_{inc}(H)$, where $M_{inc}(H)$ and $M_{dec}(H)$ are the values of magnetization with increasing and decreasing external magnetic field H, correspondingly. The so-called Bean formula derived from the critical state model [3] sets the ratio between $J_c(H)$ и $\Delta M(H)$. For an infinitely long parallelepiped, it can be represented as

$$J_c(H) = \frac{2\Delta M(H)}{a\left(1 - \frac{a}{3b}\right)}, \qquad (1)$$

where *a* and *b* are the lengths of the sides of the parallelepiped, $a \leq b$.

For an infinitely long cylinder, the formula has the form

$$J_c(H) = \frac{3\Delta M(H)}{D}, \qquad (2)$$

where D is the cylinder diameter. For samples with a significant demagnetization factor, the Bean formula provides acceptable estimates of the critical current density only for the loops measured to large values of the external field. [4].

For polycrystalline HTSC, the contribution to magnetization from intergrain currents is noticeable only in fields less than 0.01 T [5]. In this range of external fields, the ratio between $J_c(H)$ and $\Delta M(H)$ also depends on the demagnetization factor, which is determined by the shape and size of the sample and grains [6–8]. Due to the demagnetization factor of the grains and the magnetic flux trapped in the grains, an effective field induced at the grain boundaries is significantly greater than external magnetic field [9,10]. When the external field exceeds 0.1 T, the effective field suppresses intergrain currents, and the magnetization is determined only by the circulation of currents within the grains [11]. To determine the intragrain critical current density of polycrystalline HTSCs, formula (2) is used, in which the average grain size is used instead of the sample size (see, for example, [11–13]).

Thus, the magnetization of a superconductor is proportional to the magnetization of grains. Large grains should contribute most to the magnetization. If the distribution of grain sizes in the sample is sufficiently wide, using the average grain size $<D>$ in formula (2) yields an overestimate of $J_c$ due to underestimation of the contribution from larger grains. In this case, an effective size $D^*$, which is larger than the average size $<D>$, should be used for the ratio between $J_c(H)$ and $\Delta M(H)$.

In [14,15], the grain size distribution was taken into account when calculating the magnetic hysteresis loops of superconductors. Unfortunately, these studies did not consider the effect of the distribution parameters on the effective size required to determine $J_c(H)$ using the Bean formula.

In this study, we set out to establish a relationship between the parameters of grain size distribution and effective size in order to accurately assess critical current density from magnetic measurements. We performed modeling of magnetic hysteresis loops for polycrystalline superconductors with different grain size distributions, and derived analytical expressions of the effective size for these cases.

## 2. Method

The magnetization of a granular sample is defined as

$$M(H) = \frac{\sum_i M_g(H,D_i)V_g(D_i)}{\sum_i V_g(D_i)}. \qquad (3)$$

In this formula, summation is carried out over all grains. $D_i$ is the i-th grain diameter, $V_g$ – its value, and $M_g$ is its magnetization in the H field. The transition from summation to integration gives the following expression:

$$M(H) = \frac{\int_0^{D_s} f(D) M_g(H,D) V_g(D) dD}{\int_0^{D_s} f(D) V_g(D) dD}, \qquad (4)$$

where f(D) is grain size distribution function. Integration is carried out to the maximum size – the size of sample $D_s$.

The size of grains in polycrystalline samples is usually described by a lognormal distribution:

$$f(D) = \frac{1}{\sqrt{2\pi}\sigma D} e^{-\frac{(\ln(D/\lambda))^2}{2\sigma^2}}. \qquad (5)$$

The Weibull distribution is also used for ground powders:

$$f(D) = \frac{1}{\sigma\lambda} \left(\frac{D}{\lambda}\right)^{1/\sigma - 1} e^{-\left(\frac{D}{\lambda}\right)^{1/\sigma}}. \qquad (6)$$

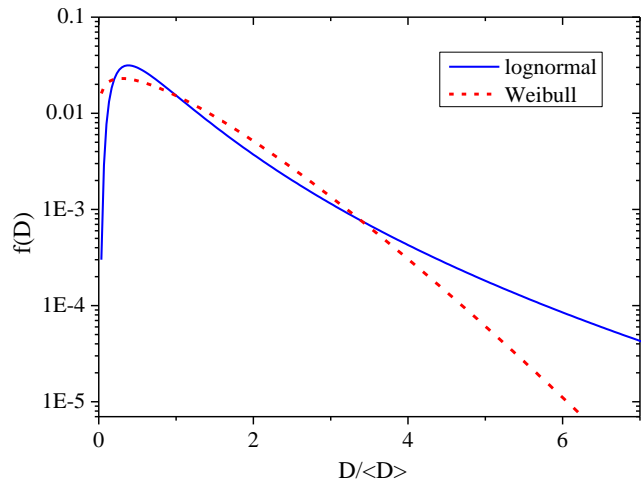

Fig.1. Log-normal distribution and Weibull distribution at σ = 0.8. The grain sizes are normalized to the average size.

In the distributions presented above, the dimensionless parameter σ sets the standard deviation of the grains, and λ is the scale parameter. The expressions for the average value <D> of the random variable in these distributions are shown in Table 1. Figure 1 shows examples of distributions (4) and (5), with σ = 0.8, plotted in semilogarithmic coordinates.

Table 1. The average size <D> and the effective size D*, expressed in the parameters of the distributions under consideration

| Distribution | ⟨D⟩ | D* |
|---|---|---|
| Log-normal | $\lambda \exp(\sigma^2/2)$ | $\langle D \rangle \exp(2\sigma^2)$ |
| Weibull | $\lambda \Gamma(1+\sigma)$, Γ – gamma function | $\dfrac{\langle D \rangle \Gamma(1+3\sigma)}{\Gamma(1+2\sigma)\Gamma(1+\sigma)}$ |

Two distributions under consideration are qualitatively similar to each other. Quantitative differences are manifested in the position of the median value (the peak of the distribution), and the different decrease rates of functions f(D) with increasing grains size. The number of large grains (with D greater than 4<D>) in the Weibull distribution is less than in the lognormal distribution.

The magnetization of a single grain is determined as the magnetization of a long superconducting cylinder with a diameter of D in an external field applied parallel to the main axis of the cylinder:

$$M_g(H,D) = -H + \frac{8}{\mu_0 D^2} \int_0^{D/2} rB\,dr, \quad (7)$$

$$\frac{dB}{dr} = \pm \mu_0 J_c. \quad (8)$$

The method used is described in detail in [15,16]. This method was successfully used to describe the magnetic hysteresis loops of various polycrystalline superconductors [17–19].

### 3. Results and discussion

Magnetic hysteresis loops were modeled for distributions (5) and (6) with different σ values. The average <D> value was the same for all the loops. The computed magnetic hysteresis loops (Figure 2) demonstrate that an increase in the σ parameter at a constant <D> value leads to an increase in ΔM. For σ ≤ 0.4, the loops, calculated using the lognormal distribution, almost coincide with those calculated using the Weibull distribution. However, for σ > 0.4, the ΔM values are higher for the lognormal distribution compared to the Weibull one.

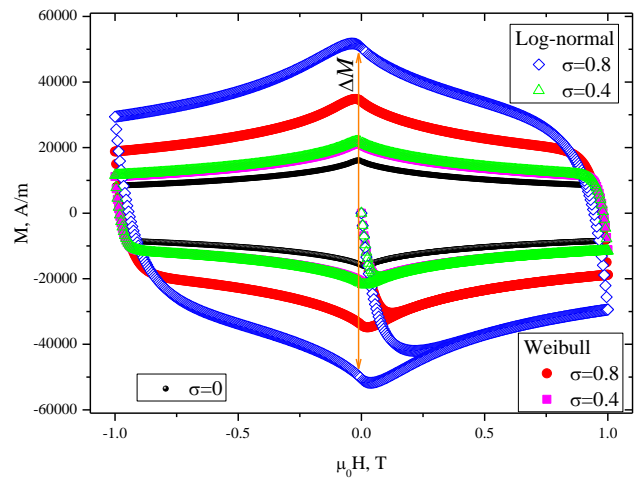

Fig. 2. Magnetic hysteresis loops with the same average grain size but different distributions.

The critical current density is the same for all grains. Therefore, the observed change in ΔM is caused by a change in the effective size $D^*$. The values of $D^* = 3\Delta M/J_c$ were determined from the obtained loops (Fig. 3) for the lognormal distribution and the Weibull distribution at different σ values. At σ > 0.2, the deviation of the ratio $D^*/\langle D \rangle$ from 1 becomes noticeable. At σ ≤ 0.4, the values of $D^*$ are the same for both distributions, and at σ ≥ 0.4, the value of $D^*$ is greater for the Weibull distribution than for the lognornal distribution. For the lognormal distribution, $D^*$ is twice the $\langle D \rangle$ at σ ≈ 0.6, as for Weibull distribution, $D^*$ is twice $\langle D \rangle$ at σ ≈ 0.7.

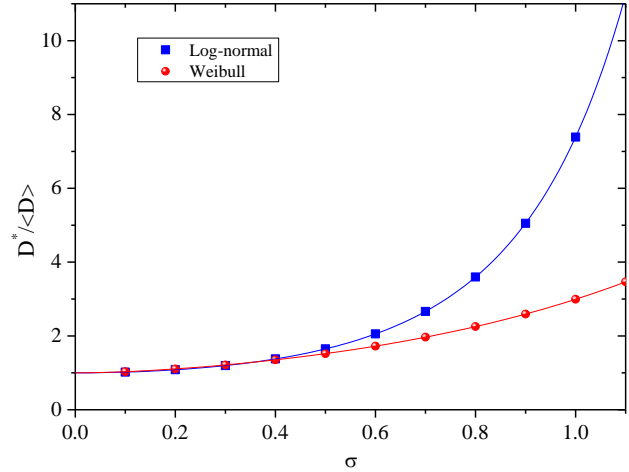

Fig. 3. Dependence of the effective size $D^*$ on the parameter σ.

In order to find analytical expressions for the dependencies of $D^*$ on σ, let's rewrite expression (4) as follows:

$$\Delta M(H) = \frac{\int_0^{D_s} f(D) \Delta M_g(H,D) V_g(D) dD}{\int_0^{D_s} f(D) V_g(D) dD}. \qquad (9)$$

The values of $\Delta M_g$ for one grain are proportional to the size: $\Delta M_g = J_c D/3$, and the volume of $V_g$ is proportional to $D^2$, since the model considers grains in the shape of a cylinder: $V_g = \pi D^2 L/4$, where L is the length of the grains. From (2) and (9) we obtain the equation for the effective size:

$$D^* = \frac{3\Delta M(H)}{J_c(H)} = \int_0^{D_s} D^3 f(D) dD \Big/ \int_0^{D_s} D^2 f(D) dD. \qquad (10)$$

By calculating (10) for distributions (5) and (6), we obtained analytical expressions for $D^*$, which are shown in Table 1. The values of $D^*$, represented by dots in Fig. 3, were obtained by fitting Equation (2) and lie on the curves described by dependencies of $D^*$ on σ (solid lines in Fig. 3). Therefore, the analytical expressions we found for $D^*$ successfully describe the relationship between $\Delta M$ and $J_c$ for different grain size distributions.

In high-temperature superconducting polycrystals produced by solid-phase synthesis, the typical grain spread corresponds to σ ≈ 0.4 ± 0.2. However, in recent work on synthesizing samples with larger grains [20], it was estimated σ ≈ 1.1. For the last case the account of the effective size $D^*$ is essential.

Comparing the critical current densities of different polycrystalline samples requires a thorough analysis of the grain size distribution in these samples, determining the type of distribution and its parameters. This analysis is quite laborious. To determine the effective size of $D^*$, required to find $J_c$, the method proposed in [21] can also be used. This method is based on the fact that asymmetry of the magnetic hysteresis loop (the module of $M_{inc}$, the magnetization values at increasing fields, is greater than the module of $M_{dec}$, the magnetization values at decreasing fields) is associated with the equilibrium magnetization of the surface layer where Abrikosov vortices are not fixed.

The thickness of the surface layer is equal to the penetration depth of the magnetic field into the superconductor $\lambda$. The observed asymmetry of the magnetic hysteresis loop allows one to determine the scale of current circulation using the expression $D^* \approx 2\lambda/[1 - (\Delta M(H_m)/2|M_m|)^{1/3}]$, where $H_m$ is the position of the minimum on the magnetization curve and $M_m$ is the minimum value of magnetization. Note that equilibrium magnetization was not taken into account in the magnetic hysteresis loops calculated in this work. However, verification showed that the influence of the parameters of grain size distribution on the values of $\Delta M$ for asymmetric and symmetric loops was similar.

## 4. Conclusion

The largest grains contribute most to the magnetization of a polycrystalline superconductor. Therefore, the values of $\Delta M$ in magnetic hysteresis loops are determined not only by the critical current density and the average size of grains, but also by a parameter $\sigma$ that characterizes the distribution of grain sizes. The effective size $D^*$, used in the Bean equation to calculate critical density, is close to the average grain size only when there is a narrow distribution ($\sigma \leq 0.2$), whereas when the distribution is wide ($\sigma > 0.2$), $D^*$ becomes much larger than the average size. It is important to consider this when extracting critical currents from magnetization curves.

When this articles has been finished, we have found that the problem of effective size for the grain size distribution was also considered in recent works [22,23].